\documentclass[%
twocolumn,
 amsmath,amssymb,
 aps,
prb,
]{revtex4-1}
\usepackage{graphicx}
\usepackage{dcolumn}
\usepackage{bm}

\begin{document}

\title{Evolution of the filamentary 3-Kelvin phase in Pb-Ru-Sr$_2$RuO$_4$ Josephson junctions}
\author{Hirono Kaneyasu$^{1}$, Sarah B. \textsc{Etter}$^{2}$, Toru \textsc{Sakai}$^{1,3}$ and Manfred \textsc{Sigrist}$^{2}$}
\affiliation{$^{1}$Department of Material Science, University of Hyogo, Kamigori, Ako, Hyogo 678-1297, Japan \\
$^{2}$Institute for Theoretical Physics, ETH Zurich, Zurich CH-8093, Switzerland \\
$^{3}$Japan Atomic Energy Agency, SPring-8, Sayo, Hyogo 679-5148, Japan}

\date{\today}

\begin{abstract}
The evolution of the filamentary 3-Kelvin (3K) superconducting phase at the interface between Sr$_2$RuO$_4$ and Ru-metal inclusions is discussed for Pb-Ru-Sr$_2$RuO$_4$ contacts. Using the Ginzburg--Landau model, the influence of proximity-induced superconductivity in Ru on the topology of the 3K phase is analyzed. Because the $s$-wave order parameter in Ru favors a 3K state of trivial topology, the onset temperature of the phase with a non-trivial topology, which is compatible with the bulk phase of Sr$_2$RuO$_4$, is essentially reduced to the bulk transition temperature. Because the topology of the superconducting state in Sr$_2$RuO$_4$ is crucial for the Josephson effect through Pb-Ru-Sr$_2$RuO$_4$ contacts, this model qualitatively reproduces the experimental observation of the anomalous temperature dependence on the critical current.
\begin{description}
\item[PACS numbers] 74.20.De,74.45.+c,74.70.Pq, 74.25.Dw
\end{description}
\end{abstract}
\maketitle

Besides the intriguing superconducting phase appearing in the quasi-two-dimensional strongly correlated metal Sr$_2$RuO$_4$ (SRO) below its bulk transition temperature $T_{c, \rm SRO}=1.5$~K \cite{Mackenzie-Modphys,Maeno-Nature,Maeno-PhysToday,Maeno-JPSJ}, the filamentary superconductivity nucleating at $ T^* \approx 3$~K in eutectic Ru-SRO samples bears further fascinating features \cite{Maeno-PRL,Yaguchi-2003,Matsumoto-2003,Kawamura-JPSJ,Yaguchi-JPSJ,Mao-PRB,Nakamura-PRB,Nakamura-JPSJ}. 
For the bulk state of SRO in the zero-magnetic field, multiple studies \cite{Mackenzie-Modphys,Maeno-Nature,Maeno-PhysToday,Maeno-JPSJ}, in particular, the observation of the polar Kerr effect \cite{Kerr} and intrinsic magnetism in $\mu$SR experiments  \cite{Musr}, count as evidence for the realization of a time-reversal symmetry breaking (TRSB) chiral $p$-wave state \cite{Mackenzie-Modphys, Maeno-Nature,Maeno-PhysToday,Maeno-JPSJ}. On the other hand, several experimental attempts to observe spontaneous edge currents expected for the chiral p-wave state led to negative results \cite{Edge_Ex1,Edge_Ex2, Edge_Theo1} and have triggered several theoretical studies exploring potential reasons for this conflicting result \cite{Edge_Theo2,Edge_Theo3,Edge_Theo4}. 
Microscopic calculations concerning the pairing symmetry show a close competition between a chiral and helical p-wave state, the former having inplane- and the latter c-axis equal-spin pairing \cite{RG}. Both of these phase are compatible with NMR-data, if we assume that the pinning of the spin configuration by spin-orbit coupling is weak \cite{NMR-1,NMR-2}. 
Also the observation of half-flux-quantum vortices \cite{HQV} is probably most easily explained with an almost freely twistable d-vector.
On the other hand, recent functional renormalization group studies support the spin triplet pairing dominantly in the $ \gamma $-band which favors the chiral p-wave channel due to spin-orbit coupling \cite{RG,FRG,NG}. In the following we will assume that the bulk superconducting phase of
SRO has the chiral p-wave symmetry. 

In eutectic systems, where excess Ru segregates from bulk SRO into micrometer-sized Ru-metal inclusions, superconductivity is believed to appear first at the interfaces between Ru and SRO at temperatures as high as $ T^* \approx 3$~K \cite{Maeno-PRL,Sigrist-JPSJ,footnote}.
This so-called ''3-Kelvin'' (3K) phase evolves into the bulk phase when the temperature is reduced. However, because the phase nucleating at $T^*$ does not break the time-reversal symmetry, the transition from the filamentary to the bulk phase involves an additional phase transition \cite{Sigrist-JPSJ,Kaneyasu-3K-JPSJ}. From tunneling spectroscopy results and the behavior of the critical current with the 3K phase, we find evidence for this additional transition at $ T' \approx 2.4 $~K \cite{Kawamura-JPSJ,Yaguchi-JPSJ,Mao-PRB}.

\begin{figure}
\begin{center}
\includegraphics[width=0.45\textwidth]{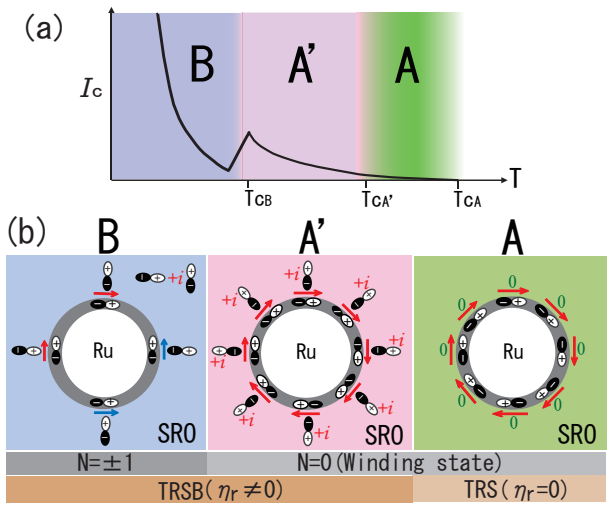}
\end{center}
\caption{(a) Schematic view of the anomalous temperature dependence of the Josephson critical current $ I_c $ in the Pb-Ru-Sr$_2$RuO$_4$ device, indicating the superconducting phase present at the Ru-Sr$_2$RuO$_4$ interface \cite{Nakamura-PRB, Nakamura-JPSJ}. In our model, the drop of $ I_c $ around $ T_c $ is connected with the transition from the $A'$- to the $ B$-phase. (b) The three filamentary phases, $ A $, $ A'$, and $ B$, are characterized by the arrangement of the order parameter components and the
phase winding number $ N $ (see text). In the $A$-phase, only $ |\eta_{\theta}| \neq 0 $ has TRS; in the $A' $-phase, both $ |\eta_{\theta}| $ and $  |\eta_{r}| \neq 0 $ have TRSB; and in the $B$-phase, both $  |\eta_{\theta}| $ and $ |\eta_{r} |\neq 0 $ have TRSB and have the same topology as the bulk phase $ k_x +i k_y $. Note the $ A $- and $ A$-phase are topologically trivial as they are invariant under rotation around the circle center, while the $ B$-phase is topologically non-trivial by acquiring a phase $ e^{i \theta} $ upon a rotation by an angle $ \theta $.}
\label{Fig-State}
\end{figure}

Our present study is motivated by experiments on a Josephson device consisting of a Pb film on top of the $c$-axis oriented surface of SRO,
yielding Pb-Ru-SRO contact through Ru inclusions. In this device geometry direct Josephson coupling between Pb and SRO is suppressed for a chiral $p$-wave state and emphasizes the path through Ru inclusions \cite{Nakamura-PRB, Nakamura-JPSJ}.
In some experimental setups, coupling through a single Ru inclusion has been achieved. In this superconducting-normal-superconducting (SNS) contact, the $s$-wave superconductivity penetrates from Pb to Ru by proximity effect. Note that Ru is a conventional superconductor with $ T_{c, \rm Ru} \approx 0.5 $~K.  The Josephson coupling between Pb and SRO appears with the onset of the 3K phase above $T_{c, \rm SRO} $ ($ T_{c, \rm Pb} = 7.2 $ K), displaying an anomalous temperature dependence of the critical current $ I_c(T) $, as shown schematically in Fig.~\ref{Fig-State} \cite{Nakamura-PRB, Nakamura-JPSJ}. The critical current, which increases with decreasing temperature within the 3K phase ($ T_{c, \rm SRO} < T < T^*$),  is interrupted by an abrupt drop
of $ I_c $ around $ T \approx T_{c, \rm SRO} $. Upon lowering the temperature further, $ I_c $ quickly recovers. It is worth noting here, that the observation of a non-vanishing Josephson effect above $ T_{c, \rm SRO} $ demonstrates the presence of the 3K phase at the interface. 

It has been suggested that the anomalous temperature originates from a change of the limiting of the Josephson current between 
SRO and Ru (with proximity-induced $s$-wave pairing) \cite{Nakamura-JPSJ, Nakamura-PRB, Sarah}. While for the temperatures above $ T_{c, \rm SRO} $ the Josephson coupling could be as essentially uniform, below $ T_{c, \rm SRO} $ the chiral nature of the bulk superconducting state in SRO would yield a frustrated Josephson coupling with a considerably lower critical temperature. 
This scenario requires that the superconducting phase of SRO at the interface switches its topology for $ T \approx T_{c, \rm SRO} $.
Here, we investigate the influence of the proximity-induced $s$-wave order parameter in Ru on the evolution of the 3K phase between $ T^* $ and $ T_{c, \rm SRO} $. The geometry we consider has Ru-SRO interfaces with normal vector in the $x$-$y$-plane for which we can expect a sizable Josephson coupling for the chiral but not for the helical $p$-wave state, since specific selection rules including spin-orbit coupling predict only lowest order Cooper pair tunneling for the former one \cite{Geshkenbein-JETP}.
Therefore, in our theoretical model, we assume the bulk chiral $p$-wave state represent in $d$-vector notation given by $ \bm{d}(\bm{k}) = \eta_0 \bm{\hat{z}} (k_x \pm i k_y) $, corresponding to a spin-triplet pairing state with angular momentum along the $z$-axis, $ L_z = \pm 1 $, and in-plane equal-spin configuration, $ S_z =0 $. 
This state is represented by a two-component order parameter through $ \bm{d}(\bm{k}) = \bm{\hat{z}} (\eta_x k_x + \eta_y k_y ) = \bm{\hat{z}} \bm{\eta} \cdot \bm{k} $ with $ \bm{\eta} = (\eta_x,\eta_y) = \eta_0 (1, \pm i) $ \cite{Sigrist-JPSJ}.
To discuss the superconducting phase at the Ru-SRO interface, it is convenient to parametrize the order parameter in terms of the parallel, $\eta_{\parallel}=\bm{\hat{z}} \cdot [ \bm{n} \times \bm{\eta} ]   $, and perpendicular, $ \eta_{\perp} = \bm{n} \cdot \bm{\eta} $, components using the interface unit normal vector $ \bm{n} $ \cite{Kaneyasu-3K-JPSJ}.

Experimental evidence suggests that the 3K phase arises from a local $T_c$ enhancement on the SRO side of the Ru-SRO interface\cite{Sigrist-JPSJ}. However, the origin is unclear. The behavior of the upper critical field suggests that the enhanced $ T_c $ region is rather narrow at 20 nm \cite{Matsumoto-2003}. Under these circumstances, only the order parameter component parallel to the interface nucleates at $ T^* $, i.e., $\eta_{\parallel}  $ \cite{Sigrist-JPSJ}. This so-called $ A$-phase is time-reversal symmetric (TRS) \cite{Kaneyasu-3K-JPSJ}. We define the phase $ \varphi(\bm{n}) = {\rm arg}( \eta_{\parallel} ) $
and its winding number  $N = \oint d\bm{s}_n \cdot \bm{\nabla}\varphi(\bm{n})$ on a path around a closed interface in the $x$-$y$-plane. Here, $ N $ characterizes the interface state topology. For the $A$-phase, $ \varphi(\bm{n}) $ is a constant such that $ N=0 $.

Three relevant interface states have been identified through their symmetry and topology: the $A$-, $A'$- and $ B$-phase states\cite{Kaneyasu-3K-JPSJ}. The latter two phases involve the appearance of the perpendicular order parameter component $ \eta_{\perp} $, which leads to TRSB. While the $A'$-phase retains the winding number $ N=0 $ (topologically trivial), the $B$-phase shows non-trivial topology with $ N = \pm 1 $, equivalent to $ \varphi(\bm{n}) = \theta_{n} = {\rm \arctan}(n_y/n_x) $. The $B$-phase is compatible with the chiral $p$-wave state. These three states are schematically depicted in Fig.~\ref{Fig-State}.
The sequence of phases from the onset of the 3K phase at $ T^*$ to the bulk superconducting state of SRO at $ T_{c, \rm SRO} $ can be discussed with a Ginzburg--Landau model, as shown in Ref. \onlinecite{Kaneyasu-3K-JPSJ}. At $ T^* \approx 3 $~K, the $A$-phase nucleates, while at the lower temperature $T'$, the $B$-phase appears through a first-order transition, simultaneously breaking TRS and switching the topology $ N=0 \to N= \pm 1 $. Note that the experimental evidence suggests $ T'_{\rm exp}  \approx 2.4 $~K\cite{Mao-PRB,Kawamura-JPSJ}. Within the $B$-phase, only percolation is required to establish bulk superconductivity at $ T_{c, \rm SRO} $ because both phases have the same topology.

The experimentally observed anomalous drop of $I_c $ at  $ T_{c, \rm SRO} $\cite{Nakamura-PRB,Nakamura-JPSJ}  can be attributed to a topology change of the superconducting state at the interface. In recent theoretical studies, the influence of the interface state topology on the critical Josephson current has been discussed assuming an $s$-wave order parameter within Ru inclusions \cite{Sarah,Kaneyasu-Vortex-JPSJ}. When $N=0$ ($A$- and $A'$-phase), the interface acts like an ordinary extended Josephson junction \cite{Owen}, while $ N=\pm 1$ ($B$-phase) yields a frustrated junction\cite{Kaneyasu-Vortex-JPSJ}. This frustration leads to a spontaneous magnetic flux pattern on the interface. In this case, the supercurrent through the interface is limited by a magnetic flux pinning-depinning transition, leading to a reduction of the Josephson critical current. This feature could explain the observed anomaly if the change of the winding number  $ N $ occurs around $ T_{c, \rm SRO}$.

For the bare Ru-SRO eutectic samples, the change of $ N $ is associated with a first-order phase transition at $ T' \approx 2.4 $~K \cite{Kaneyasu-3K-JPSJ}. If Ru hosts an $s$-wave order parameter,
we expect that the phase diagram within the 3K phase ($ T_{c, \rm SRO} < T < T^* $) will be modified through the coupling of this order parameter to the interface state. We consider the rather simple model geometry of a cylindrical Ru-metal inclusion with radius $R$, whose axis is oriented along the $z$-axis of SRO. We use a Ginzburg--Landau model for the two-component $p$-wave superconductor with the order parameter $ \bm{\eta} = (\eta_x , \eta_y) $ in the region of SRO outside the Ru inclusion. We consider two spatial dimensions perpendicular to $z$ and assume homogeneity along the $z$-axis.
The $s$-wave superconductivity of the Ru inclusion enters via interface coupling. The Ginzburg--Landau free-energy functional is then a scalar under all symmetries of SRO and consists of the SRO bulk terms\cite{Sigrist-RevModPhys, Heeb-PRB} and the interface terms at $r=R$ \cite{Kaneyasu-3K-JPSJ}:
\begin{align}
\mathcal{F} =& \int_{r > R} d^2r \Big[ a(r) | \bm{\eta} |^2 + \frac{1}{4} b \left\{ | \bm{\eta}|^4 + 2 | \eta_+|^2| \eta_-|^2 \right\}  \nonumber\\
 &+ \frac{1}{2} K \left\{ | \bm{D} \eta_+ |^2 + | \bm{D} \eta_-|^2 +\frac{1}{2} ( (D_+ \eta_-)^* (D_- \eta_+) + \mathrm{c.c.} ) \right\}\nonumber\\
  &+ \frac{(\bm{\nabla} \times \bm{A})^2}{8\pi} \Big] \nonumber\\
  &+ R \int_{int} d\theta  \left\{
   K_r | \bm{n} \cdot \bm{\eta} |^2
+ K_{\theta} |\bm{\hat{z}} \cdot (\bm{n} \times \bm{\eta}) -  \psi_s |^2 \right\},
\label{free energy}
\end{align}
with the covariant gradient defined as  $ \bm{D} = \bm{\nabla} - i \gamma\bm A $, where $\gamma= 2 \pi/ \Phi_0$ (flux quantum $ \Phi_0 = hc/2e $). We define $ \eta_{\pm} = (\eta_x \pm i \eta_y)/\sqrt{2} $ and $ D_{\pm} = D_x \pm i D_y $. The enhancement of the critical temperature near the interface is considered by defining $ a(r) = a'(T-T_c(r))$  for an $r$-dependent critical temperature $ T_c(r) = T_{c} + T_0/ \cosh[(r-R)/d] $ with $  r \geq R $, where $ d $ describes the extension of the region of the enhanced critical temperature and $ T_c = T_{c, \rm SRO} $. The bulk parameters $a'$, $b$, and $K$ are determined through the homogeneous and the linearized Ginzburg--Landau equations via the ratios $a'/b=2$ and $K/a'T_c=\xi_0^2$, where $\xi_0$ is the zero-temperature coherence length. The first integral is taken over the space outside the the Ru cylinder. The second integral over the interface between Ru and SRO describes the influence of the interface, including the coupling between $ \bm{\eta} $ and the $s$-wave order parameter $ \psi_s $ inside Ru. By symmetry and owing to spin-orbit coupling, only the order parameter component parallel to the interface, $ \eta_{\parallel} =\eta_{\theta} $, couples (with a coupling strength given by the coefficient $ K_{\theta} > 0 $ \cite{Geshkenbein-JETP} ($ \bm{n} =\hat{\bm{r}}= (\cos \theta, \sin \theta) $). The perpendicular component $\eta_{\perp}= \eta_r  $ with the coefficient $ K_r > 0 $ is reduced at the interface. The winding number $N$ enters as $\eta_{\pm}(r,\theta,z)=\eta_{\pm}(r,z)e^{i(N\pm1)\theta}$ for a single-valued order parameter.

We express $\eta_r(r)$ and $\eta_{\theta}(r)$  by $\eta_r(r) = (\eta_+ + \eta_-)/\sqrt{2}$ and $\eta_\theta(r) = -( \eta_+ - \eta_-)/\sqrt{2}$, respectively. The influence of $ \psi_s $ is evident from the interface terms in Eq. (\ref{free energy}),
\begin{align}
\mathcal{F}_{int} &= 2\pi R \Big[  K_{r}  |\eta_{r}(R)|^2 | +  K_{\theta} \{ | \eta_{\theta}(R)|^2 + |\psi_s|^2 \} \Big] \nonumber \\
&\hspace{-0.5cm} -  2 K_{\theta} \int d \theta \; | \eta_{\theta}(R) | | \psi_s| \cos (\phi - N \theta)  
\label{interface}
\end{align}
where $ \phi$ is the global phase difference between  the order parameters on the two sides of the interface.
The integral of the last term leads to $ - 2 \pi  \delta_{N,0} K_{\theta} | \eta_{\theta}(R) | | \psi_s| \cos \phi $, giving rise to the Josephson coupling only if the interface state is topologically trivial ($ N=0 $).

Obviously the winding number $ N $ is essential for the Josephson coupling. A non-vanishing $ N $ yields frustration in the phase of the two superconductors and leads to an effective decoupling \cite{Kaneyasu-3K-JPSJ}. Only $ N = 0 $ leads to uniform coupling over all the interface and allows minimization of the coupling energy by setting the phase $ \phi =0 $. The variational minimization including the interface terms yields the boundary conditions at $ r = R $,
\begin{align}
\left.\frac{d \eta_r}{d r} \right|_{r = R}&=\frac{1}{3} \left[ \left\{ \frac{4 K_r}{K}-\frac{1}{R} \right\} \eta_r(R)+\frac{N}{R}\eta_\theta(R) \right] \\
\left.\frac{d \eta_{\theta}}{d r} \right|_{r = R}&=\left\{ \frac{4 K_\theta}{K}+\frac{1}{R} \right\} \eta_\theta(R)-\frac{N}{R}\eta_r(R)-\frac{4 K_\theta}{K}\psi_s\delta_{N,0}
\label{boundary}
\end{align}
Note that the second equation expresses the coupling of $ \eta_\theta $ to $ \psi_s$.

The filamentary nature makes the 3K phase susceptible to the influence of the interface.
For $ |\psi_s|=0 $ we encounter the $ A $- and $B$-phases, which are separated by a first-order phase transition at $ T' $.
The coupling at the interface for finite $|\psi_s|>0$, which is strong and unfrustrated for $ N=0 $, works in favor of the phases with this topology. In this way, even the $A'$-phase may appear. The transition from $A$ to $ A'$ is a second-order transition because the topology is unchanged even though the TRS is spontaneously broken. A first-order transition from the $ A' $- to the $ B$-phase without symmetry breaking is inevitable as $ B $ corresponds to the SRO bulk superconducting phase.

To confirm this scenario, we numerically minimize the free-energy functional under the given boundary conditions using a relaxation method. Analyzing the behavior of the order parameter $ \bm{\eta} $, we determine the phase diagram for a varying $s$-wave order parameter $ \psi_s $ in Ru. While we do not intend a
full quantitative discussion, which is beyond the GL-formulation, we nevertheless choose our parameter in the absence of superconductivity in the Ru inclusion ($ |\psi_s| = 0 $) to roughly obtain the behavior observed experimentally: $ a'=1 $ and $ T_c $ for the bulk as well as for the interface $ K_r/K = 0.125$, $ K_{\theta}/K = 0.025$, $ T_0 = 3.6$ K with  $ d = 0.5 $ and $ R = 5 $ in length units of $ \xi_0 $.  For $ |\psi_s| = 0 $, this set of parameters yields the onset of the A-phase at $ T^* \approx 3 $ K and the transition to the B-phase at $ T' \approx 2.4 $  K.

\begin{figure}
\begin{center}
\includegraphics[width=0.46\textwidth]{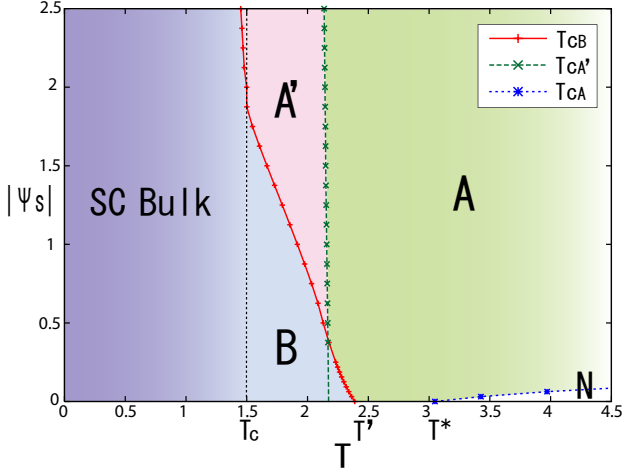}
\end{center}
\caption{Phase diagram of $ T $ versus $ |\psi_s|$. The proximity-induced order parameter $ \psi_s $ favors the topologically trivial $A'$-phase ($N=0$) relative to the topologically non-trivial $ B$-phase ($N=\pm 1$). The transition line $ T_{cA'} $ is second order and spontaneously breaks the TRS, while $ T_{cB} $ is first order and switches the topology ($ N=0 \to N=\pm 1$). Note the dashed line indicating the extension of $ T_{cA'} $ towards
$ | \psi_s| = 0 $ is covered by the $ B$-phase. The position of the intersection of $ T_{cA'} $ and $ T_{cB} $ depends on model parameters. }
\label{Fig-Phase}
\end{figure}

\begin{figure}
\begin{center}
\includegraphics[width=0.49\textwidth]{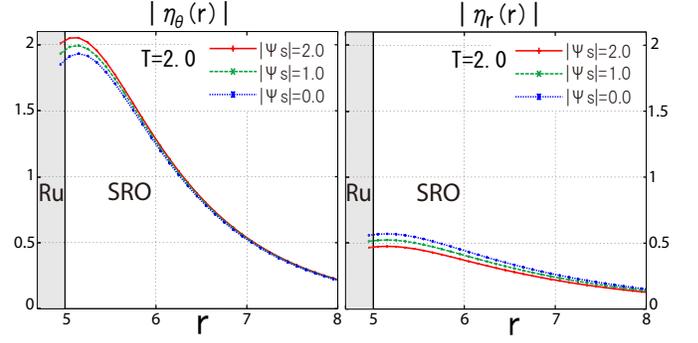}
\end{center}
\caption{Numerically calculated order parameter components of the filamentary phase at $ T = 2.0 $K for different values of $ |\psi_s| $. The azimuthal component $\eta_\theta $ couples directly to $ \psi_s $ (Eq.\ref{interface}) and grows with increasing $|\psi_s|$, while the radial component $ \eta_r $ simultaneously decreases due to competition with $ \eta_{\theta} $ and the lack of coupling to $ \psi_s $ by symmetry. }
\label{Fig-Eta-Psi}
\end{figure}

\begin{figure}
\begin{center}
\includegraphics[width=0.48\textwidth]{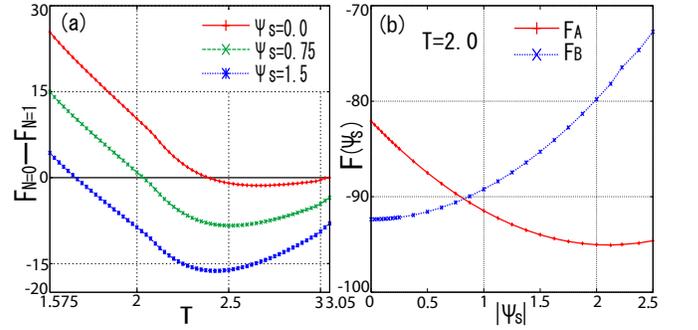}
\end{center}
\caption{Comparison of the free energy for the two topological sectors: (a) Free-energy difference between the phases with $ N=0 $ and $ N=1 $. The zero-crossing marks the first-order transition $ T_{cB} $. The soft anomaly in the free-energy difference around $ T = 2.2 $ K corresponds to the second-order transition at $ T_{cA'} $. (b) Free energy of the $A/A'$- and $ B$-phase, given by $ F_A $ and $F_B$, respectively, as a function of $ |\psi_s| $ for $ T = 2.0 $~K$ < T' $. While the $ A/A'$-phase lowers its energy (interface energy), the $ B$-phase becomes energetically unfavorable with increasing $ |\psi_s| $ with a crossing of the two free energies at the finite value of $| \psi_s| $ (first-order transition).}
\label{Fig-Tb}
\end{figure}

We summarize our results in a phase diagram of temperature $T$ versus $s$-wave order parameter strength $|\psi_s|$  (see Fig.~\ref{Fig-Phase}). 
The onset of the 3K phase is shifted to higher temperatures as the coupling between $ \eta_{\theta} $ and $ \psi_s $ (Eq.\ref{interface}) facilitates the nucleation of the $A$-phase whose winding $ N=0 $ is compatible with the uniform $ \psi_s $.
This is underlined by the behavior observed in Fig.~\ref{Fig-Eta-Psi}(a), where increasing $ |\psi_s| $ strengthens the component $ \eta_{\theta} $ while $ \eta_r $ (yielding TRSB) is slightly reduced (Fig.~\ref{Fig-Eta-Psi}(b)), which results from the competition between the two order parameter components \cite{Matsumoto-1999}. Note that both order parameter components in SRO, $ \eta_{\theta} $ and $ \eta_r $,  decay on the length scale of the coherence length away from the interface, featuring a filamentary phase (Fig.\ref{Fig-Eta-Psi}). 
Upon increasing $ | \psi_s| $, the transition to the $B$-phase, changing $N$ from 0 to $ \pm 1 $, is shifted to lower temperatures until the transition temperature $ T_{cB} $ approaches the bulk critical temperature $ T_{c} = 1.5 $ K. This shift is a result of the competition with the topologically trivial phase, $A$ and $A'$ ($N=0$). 
Indeed, for large enough $ | \psi_s| $, the  $A'$-phase appears in the temperature range otherwise covered by the $ B$-phase. 
In contrast the $T_{cB}$ shifts, the line of $T_{cA'}$ (yielding TRSB) is rigid with increasing $ |\psi_s|$.
Considering the free energies of the topological sectors $ N=0 $ and $ N= \pm 1 $, we find a first-order transition for $ T_{cB} $, where $ F_{N=0} = F_{N=1} $ (see Fig.~\ref{Fig-Tb}(a)). The effect of the competition between the two topological sectors is also obvious from the comparison between $ F_{N=0} $ and $ F_{N=1} $ in Fig.\ref{Fig-Tb}(b). The Josephson coupling (Eq.\ref{interface}) reduces the free energy for the phases $A$ and $A'$ ($N=0$) with increasing $ |\psi_s| $, while the interface energy only increases ($ \propto |\psi_s|^2 $) for the $B$-phase, which has no coupling to $ \psi_s $.
The transition between $ A $ and $A'$ at $ T_{cA'} $ continuously yields spontaneous TRS breaking through the appearance of $ \eta_r $. Eventually, $ T_{cB} $ reaches
$ T_{c, \rm SRO} $ at a finite value $ |\psi_s| \approx 1.8 $ in our units. For larger $ |\psi_s| $, the $ A'$-phase seems to extend into the temperature range where the bulk phase of SRO is already present, which is, however, a numerical artifact due to the finite size $R_{disk}$ of the model system used for computation.

We now analyze our result with regard to the Josephson effect in the Pb-Ru-Sr$_2$RuO$_4$ device. Within SRO, the coherence length along the $z$-axis is roughly 20 times shorter than that in the $x$-$y$-plane, so we may consider the RuO$_2$-planes to be weakly coupled. The proximity-induced $s$-wave order parameter $\psi_s$ in Ru decreases quickly with distance from the Pb-Ru contact. Thus, the nearly independent layers of SRO are exposed to a different magnitude of $|\psi_s|$ for the corresponding $z$-coordinate.  The $ A $- and $A'$-phases dominate the Josephson effect above $ T_{c, \rm SRO} $, and a finite critical current becomes observable when the $A$-phase reaches a sufficient magnitude. This junction is unfrustrated ($N=0$) and shows a monotonically increasing critical current with decreasing temperature until $ T_{cB} $ is reached.
The overall Josephson current is obviously dominated by the region very close to the Pb-Ru contact. Here, we may expect that the
onset of the $B$-phase, $ T_{cB} $,  is shifted to $T_{c, \rm SRO}$. At $ T_{cB} $, the topology of the Josephson contact changes. Because of the change from $ N=0 $ to $ N= \pm 1 $, the Josephson critical current $I_c$ is reduced\cite{Sarah}. An anomalous drop should occur near $T_{c, \rm SRO}$, followed by an increase at lower temperatures, as shown in Fig.~\ref{Fig-State}. Because the order parameter at the interface retains its  $ A'$-phase character until the bulk transition, it may change through a domain-wall-like twist into the bulk phase at temperatures below $T_{c, \rm SRO}$. However, this feature cannot be treated within our computational approach, which is restricted to sectors of fixed $ N $-values. In addition to the $z$-dependence of $ \psi_s$, which yields a transition spread between the topologically distinct interface states, this feature is likely responsible for the width of the $I_c$ temperature decrease.

With our simple model, we have demonstrated that the $s$-wave superconductivity induced in the Ru inclusion by proximity to the Pb contact changes the evolution of the  filamentary 3K phase to Sr$_2$RuO$_4$ bulk superconductivity, compared to the case with a metallic Ru inclusion and no Pb contact\cite{Kaneyasu-3K-JPSJ}. The experimental observation of the anomalous temperature dependence of the  critical current  in the Pb-Ru-SRO Josephson junction reflects this modification, demonstrating that the Josephson current is limited by distinct mechanisms in phases with $ N=0 $ ($A $- and $A'$-phases) and with $ N= \pm 1 $ ($B$-phase) \cite{Sarah}. The complete description of the temperature dependence of the Josephson current requires the discussion of the third spatial dimension of the device, which we avoid in this paper. Although our model is simplified, we believe that it captures the most essential features of the experimental setup. The mechanism of the 3K phase evolution in response to the Josephson coupling is consistent with that of the observed anomalous critical current behaviour, and thus it provides evidence of the existence of the 3K phase with the winding state.

We are very grateful for discussions with Y. Maeno, T. Nakamura, T. Nomura, Y. Hasegawa, F. Mila, and the late N. Hayashi. This study was financially supported by the Japan Securities Scholarship Foundation. S.~E. and M.~S. are grateful for financial support by a grant from the Swiss National Science Foundation.


\end{document}